\documentclass[reprint,amsmath,amssymb,superscriptaddress]{revtex4-1}

\usepackage{graphicx}
\usepackage{dcolumn}
\usepackage{bm}
\usepackage{color}

\begin{document}


\title{Crucial role of interfacial $s$-$d$ exchange interaction in the temperature dependence of tunnel magnetoresistance}


\author{Keisuke Masuda}
\email{MASUDA.Keisuke@nims.go.jp}
\affiliation{Research Center for Magnetic and Spintronic Materials, National Institute for Materials Science (NIMS), Tsukuba 305-0047, Japan}
\author{Terumasa Tadano}
\affiliation{Research Center for Magnetic and Spintronic Materials, National Institute for Materials Science (NIMS), Tsukuba 305-0047, Japan}
\author{Yoshio Miura}
\affiliation{Research Center for Magnetic and Spintronic Materials, National Institute for Materials Science (NIMS), Tsukuba 305-0047, Japan}
\affiliation{Center for Spintronics Research Network, Graduate School of Engineering Science, Osaka University, Toyonaka, Osaka 560-8531, Japan}


\date{\today}

\begin{abstract}
The tunnel magnetoresistance (TMR) is one of the most important spintronic phenomena but its reduction at finite temperature is a severe drawback for applications. Here, we reveal a crucial determinant of the drawback, that is, the $s$-$d$ exchange interaction between conduction $s$ and localized $d$ electrons at interfacial ferromagnetic layers. By calculating the temperature dependence of the TMR ratio in Fe/MgO/Fe(001), we show that the obtained TMR ratio significantly decreases with increasing temperature owing to the spin-flip scattering in the $\Delta_1$ state induced by the $s$-$d$ exchange interaction. The material dependence of the coupling constant $J_{sd}$ is also discussed on the basis of a nonempirical method.
\end{abstract}


\maketitle

\begin{figure}
\includegraphics[width=8.5cm]{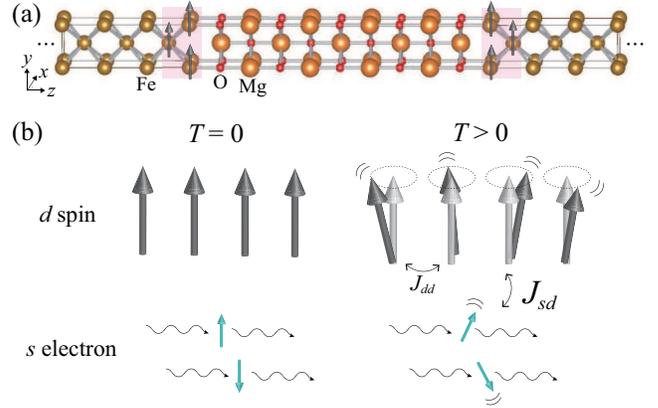}
\caption{\label{Fig1} (a) An Fe/MgO/Fe(001) MTJ. Spin fluctuations in the shaded interfacial layers provide a significant reduction of the TMR ratio with increasing temperature. (b) Illustrations of our idea. When the temperature increases, the spins of $s$ electrons fluctuate through the exchange coupling with $d$-electron spins, which reduces the TMR ratio significantly.}
\end{figure}

Understanding the physics of spin transport at finite temperature is of great importance not only from fundamental but also from application points of view. A particularly challenging issue is the temperature decay of the tunnel magnetoresistance (TMR) in magnetic tunnel junctions (MTJs) [Fig. \ref{Fig1}(a)], which are used for various magnetic sensors and nonvolatile magnetic random access memories. Although a giant TMR ratio has been demonstrated at low temperature in various MTJs \cite{2004Parkin-NatMat,2004Yuasa-NatMat,2006Sakuraba-APL,2008Tsunegi-APL,2009Ishikawa-APL,2009Tezuka-APL,2014Li-PRB,2015Liu-JPD,2016Hu-PRB,2016Moges-PRB}, its significant reduction with increasing temperature has also been observed \cite{2006Sakuraba-APL,2008Tsunegi-APL,2009Ishikawa-APL,2009Tezuka-APL,2014Li-PRB,2015Liu-JPD,2016Hu-PRB,2016Moges-PRB,2021Scheike-APL}. This is a critical problem to be solved, since MTJs are usually used at room temperature.

A clue to explain this phenomenon is conduction $sp$-electron states in ferromagnets; several experiments \cite{2014Li-PRB,2016Moges-PRB} have shown that $sp$-electron states with a smaller effective mass than $d$-electron states provide dominant contributions to transport properties of MTJs. However, most previous theories \cite{1987Liechtenstein-JMMM,2000Katsnelson-PRB,2001Pajda-PRB,2015Kvashnin-PRB,2006Lezaic-PRL,2008Chioncel-PRL,2020Shinya-APL,2020Nawa-PRB} have focused only on $d$-electron states and the $d$-$d$ exchange interaction between $d$ electrons on neighboring sites. This is because $d$-electron states have a large density of states around the Fermi level and play the main role for static magnetic properties in bulk ferromagnets at finite temperature. For example, the Curie temperatures of $3d$ transition metals have been estimated by the Heisenberg model with the $d$-$d$ exchange interaction \cite{1987Liechtenstein-JMMM,2000Katsnelson-PRB,2001Pajda-PRB,2015Kvashnin-PRB}. Moreover, the temperature dependencies of spin polarizations in Heusler alloys have been understood by spin fluctuations in $d$-electron states \cite{2006Lezaic-PRL,2008Chioncel-PRL,2020Shinya-APL,2020Nawa-PRB}. In contrast, since transport properties in MTJs can be dominated by $sp$-electron states as mentioned above, we need to clarify how these states correlate with the temperature dependence of the TMR ratio.

In this Letter, we show that an intra-atomic $s$-$d$ exchange interaction between conduction $s$ and localized $d$ electrons plays a significant role for the temperature decay of the TMR ratio in Fe/MgO/Fe(001). While this interaction plays an essential role for the well-known Kondo effect \cite{Hewson}, its importance in the TMR effect has yet to be suggested. We calculate the temperature dependence of the TMR ratio by employing the tight-binding model with the $s$-$d$ exchange interaction. As shown in Fig. \ref{Fig1}(b), increasing the temperature enhances spin fluctuations in $d$-electron states, as suggested in previous studies \cite{2006Lezaic-PRL,2008Chioncel-PRL,2020Shinya-APL,2020Nawa-PRB}. We find that such spin fluctuations propagate from $d$- to $s$-electron states through the $s$-$d$ exchange interaction. As a result, spin-flip scattering occurs in $s$-electron states, leading to a significant reduction of the TMR ratio \cite{remark_spin-flip}. These findings indicate that the $s$-$d$ not $d$-$d$ exchange interaction is the main origin of the TMR reduction, since the TMR ratio never drops significantly for a small $s$-$d$ exchange interaction even if $d$ spins fluctuate. We also find that the $s$-$d$ exchange interaction at interfacial ferromagnetic layers contributes dominantly to the TMR reduction. We will finally estimate the coupling constant of the $s$-$d$ exchange interaction using a nonempirical method. We show that the material dependence of the TMR reduction can be explained by the estimated coupling constants. Our results would be quite important for future materials design for a smaller temperature dependence of the TMR ratio.

Our calculation is based on the tight-binding model,
\begin{equation}
H_0=\sum_{ij} \sum_{\mu\nu\sigma}\, t^{\mu\nu}_{ij} c^\dagger_{i \mu \sigma}c_{j\nu\sigma} + \sum_{i\mu\sigma}\, \epsilon^\mu_{i\sigma}n_{i\mu\sigma},
\end{equation}
where $c^\dagger_{i \mu \sigma}$ creates an electron with spin $\sigma$ in orbital $\mu$ at site $i$, $t^{\mu\nu}_{ij}$ is the hopping integral of electrons, $\epsilon^\mu_{i\sigma}$ is the on-site potential measured from the Fermi level $E_{\rm F}$, and $n_{i\mu\sigma}=c^\dagger_{i\mu\sigma}c_{i\mu\sigma}$. We constructed this type of Hamiltonian for bcc Fe and MgO \cite{remark_tb-model}. In addition to the one-body terms $H_0$, we considered the $s$-$d$ exchange interaction at each Fe site,
\begin{equation}
H_{sd}=-2J_{sd}\sum_{i} {\bm s}_{i} \cdot {\bm S}_{i},\label{eq2}
\end{equation}
where ${\bm s}_{i} \equiv \frac{1}{2} \sum_{\sigma\sigma'} c^\dagger_{is\sigma} {\bm \tau}_{\sigma\sigma'} c_{is\sigma'}$ is the spin operator for $s$ electrons with ${\bm \tau}_{\sigma\sigma'}$ being the Pauli matrices and ${\bm S}_{i}$ is the operator for the localized spin in the $d$ orbitals, which is assumed to have $S=2$ because of six $3d$ valence electrons in Fe. Since the $p$-orbital states are far from the Fermi level and do not affect our results, we neglected the $p$-$d$ exchange interaction between the $p$ and $d$ electrons. When the temperature increases, the localized spin ${\bm S}_{i}$ has fluctuations in the longitudinal ($S_{iz}$) and transverse ($S_{ix}$ and $S_{iy}$) directions, leading to spin-flip scattering in the $s$-electron states through $H_{sd}$. We treat this spin-flip scattering at finite temperature by mixing the up-spin and down-spin $s$ states within the coherent potential approximation (CPA) \cite{1996Takahashi-PRB,2000Itoh-PRL}. By introducing the orbital-diagonal coherent potentials $\Sigma_{s\sigma}$ ($\sigma = \uparrow, \downarrow$) in the $s$ orbital, the Hamiltonian $H=H_0+H_{sd}$ of Fe is rewritten as  $H=K+V$, with $K=H_0 + \sum_{i\sigma}\, \raisebox{-0.35ex}[1ex][0ex]{$\Sigma_{s\sigma}$}\, c^\dagger_{is\sigma}c_{is\sigma}$ and $V=\sum_{i\sigma \sigma'}\, c^\dagger_{is\sigma}\, \left(-J_{sd}\,{\bm \tau}_{\sigma\sigma'}\cdot{\bm S}_i - \raisebox{-0.35ex}[1ex][0ex]{$\Sigma_{s\sigma}$}\,\delta_{\sigma\sigma'}\right)\, c_{is\sigma'}\equiv \sum_i v_i$. Using $v_i$ and the unperturbed Green's function $P\equiv 1/(\omega-K)$, the scattering operator $t_i$ is defined as $t_i=v_i\,(1-Pv_i)^{-1}$, where $\omega$ is the energy relative to the Fermi level and is set to 0. We determined the values of $\Sigma_{s\sigma}$ at each temperature from the CPA condition $\langle t_i \rangle =0$. Technical details for solving $\langle t_i \rangle =0$ are presented in the Supplemental Material \cite{Supplemental}. To solve this, we assumed a typical temperature dependence of $\langle S_{iz} \rangle$, $\langle S_{iz} \rangle = S \sqrt{1-\left( T/T_{\rm C} \right)^2}$, where $T_{\rm C}$ is the Curie temperature of Fe ($T_{\rm C}=1040\,\,{\rm K}$). As the temperature increases, $\langle S_{iz} \rangle$ decreases following this equation. Such a decrease in $\langle S_{iz} \rangle$, i.e., the enhancement of $d$-spin fluctuation, propagates to $s$-electron states through the $s$-$d$ exchange interaction [Eq. (\ref{eq2})], leading to the spin-flip $s$-electron scattering. The real part of $\Sigma_{s\sigma}$ gives an exchange splitting and the imaginary part gives a finite lifetime, namely, the occurrence of spin-flip scattering from $|s,\sigma\rangle$ to $|s,\bar{\sigma}\rangle$. Since the effect of the exchange splitting is already included in the on-site potential in $H_0$, we considered only the imaginary part of the coherent potential ${\rm Im}(\Sigma_{s\sigma})$ in our transport calculations.

\begin{figure}
\includegraphics[width=8.1cm]{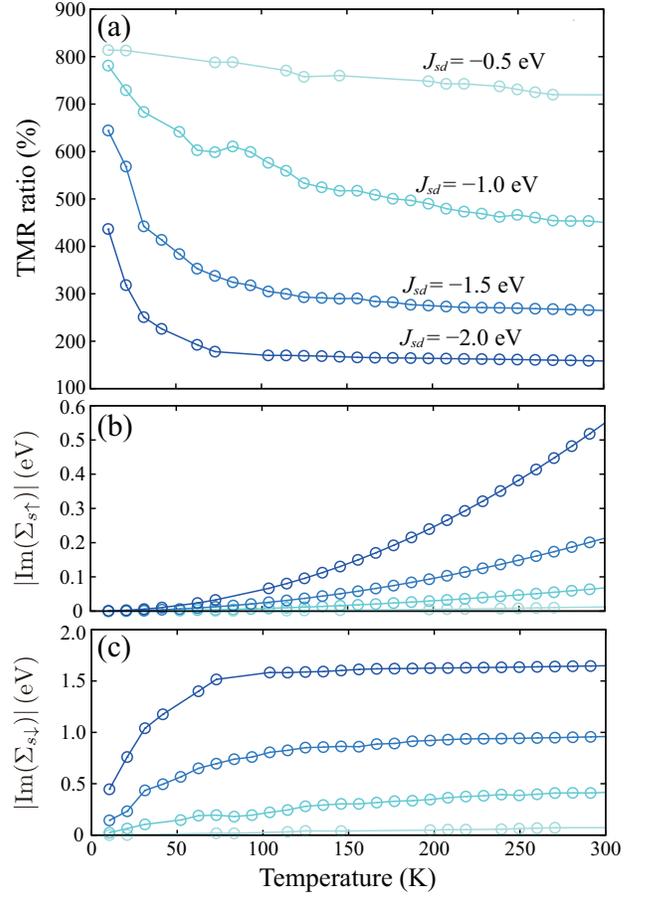}
\caption{\label{Fig2} Temperature dependencies of (a) TMR ratios, (b) $|{\rm Im}(\Sigma_{s\uparrow})|$, and (c) $|{\rm Im}(\Sigma_{s\downarrow})|$ for different values of the exchange interaction $J_{sd}$.}
\end{figure}
\begin{figure*}
\includegraphics[width=18.0cm]{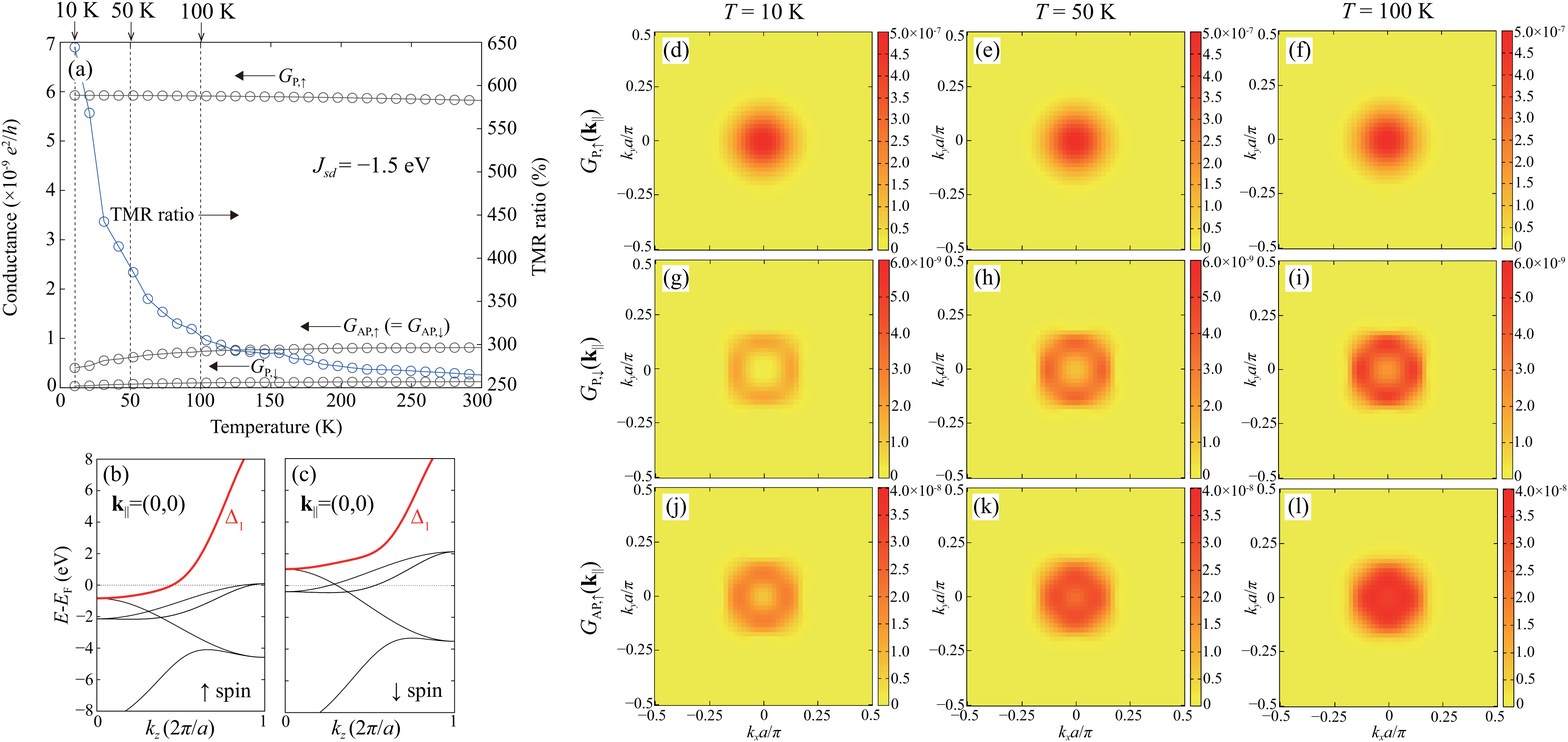}
\caption{\label{Fig3} (a) Temperature dependencies of the conductances and TMR ratio for $J_{sd}=-1.5\,{\rm eV}$. (b), (c) Up- and down-spin bands of Fe along the $\Delta$ line contributing dominantly to the TMR effect. (d)--(l) The ${\bf k}_\parallel$-resolved conductances calculated for $J_{sd}=-1.5\,{\rm eV}$ \cite{remark_kpara-dep}. (d)--(f) $G_{{\rm P},\uparrow}({\bf k}_\parallel)$ at $T=10$, $50$, and $100\,\,{\rm K}$, respectively. (g)--(i) The same as (d)--(f) but for $G_{{\rm P},\downarrow}({\bf k}_\parallel)$. (j)--(l) The same as (g)--(i) but for $G_{{\rm AP},\uparrow}({\bf k}_\parallel)$. The unit of the color bars in (d)--(l) is $e^2/h$.}
\end{figure*}
We calculated the electronic states of Fe/MgO/Fe(001) by using the recursive Green's function method \cite{1981Lee-PRL,1997Umerski-PRB} in combination with the above-mentioned parameters ($t^{\mu \nu}_{ij}$, $\epsilon^{\mu}_{i\sigma}$, and $\Sigma_{s\sigma}$). This method allows us to calculate the Green's function at the ($n$+1)th layer ${\bf g}_{n+1}$ from that at the $n$th layer ${\bf g}_n$: ${\bf g}_{n+1}=(\omega-{\bm \epsilon}-{\bf t}^{\dagger}{\bf g}_{n}{\bf t})^{-1}$. Here, ${\bm \epsilon}$ is the on-site potential matrix including $\epsilon^{\mu}_{i\sigma}$ and $\Sigma_{s\sigma}$ and ${\bf t}$ is the hopping-integral matrix composed of $t^{\mu \nu}_{ij}$. In addition to the hopping integrals of bulk Fe and MgO, we also need those at the interface, which were approximately determined by applying Harrison's method \cite{1989Harrison-Dover} to the hopping integrals of bulk Fe. Starting from the left (right) surface Green's function of Fe, we obtained the Green's function at each layer from Fe to MgO by using the above recursive equation, leading to the Green's functions of the left (right) semi-infinite system \cite{1997Umerski-PRB}. From these we can obtain the Green's functions of the entire system Fe/MgO/Fe(001) \cite{1997Mathon-PRB}. By applying the Kubo-Greenwood formula \cite{1997Mathon-PRB,1997Mathon-PRB_2} to the obtained Green's functions, temperature-dependent conductances were calculated. Since our system has translational symmetry in the $xy$ plane, the electronic states are labeled by the in-plane wave vector ${\bf k}_\parallel=(k_x,k_y)$. The conductances $G_{\rm P}({\bf k}_\parallel)=G_{{\rm P},\uparrow}({\bf k}_\parallel)+G_{{\rm P},\downarrow}({\bf k}_\parallel)$ and $G_{\rm AP}({\bf k}_\parallel)=G_{{\rm AP},\uparrow}({\bf k}_\parallel)+G_{{\rm AP},\downarrow}({\bf k}_\parallel)$ for parallel and antiparallel magnetization configurations were calculated for each ${\bf k}_\parallel$ and were averaged as $G_{\rm P}=\sum_{{\bf k}_\parallel}G_{\rm P}({\bf k}_\parallel)/N$. Here, the sampling number $N$ of ${\bf k}_\parallel$ points was set to $100 \times 100$ for ensuring good convergence of the conductances. The TMR ratio was estimated using the optimistic definition, ${\rm TMR\,\,ratio}\,(\%)=100\times(G_{\rm P}-G_{\rm AP})/G_{\rm AP}$.

Figure \ref{Fig2}(a) shows the temperature dependencies of the TMR ratio for different values of $J_{sd}$. We focused on negative values of $J_{sd}$ because they are reasonable as discussed later. The TMR ratio decreases with increasing the temperature for all the values of $J_{sd}$. When the temperature increases, the imaginary part of the coherent potential $|{\rm Im}(\Sigma_{s\sigma})|$ increases as shown in Figs. \ref{Fig2}(b) and \ref{Fig2}(c), which means an enhancement of spin-flip scattering and leads to the reduction of the TMR ratio. A larger $|J_{sd}|$ gives a faster decrease in the TMR ratio because of a faster increase in $|{\rm Im}(\Sigma_{s\sigma})|$ with increasing the temperature. More detailed behaviors of $|{\rm Im}(\Sigma_{s\sigma})|$ shown in Figs. \ref{Fig2}(b) and \ref{Fig2}(c) can be understood as follows. Note here that the $s$-$d$ exchange interaction [Eq. (\ref{eq2})] can be rewritten as $H_{sd}=-2J_{sd}\sum_{i}\,[\frac{1}{2}\,(s_{i+} S_{i-}+s_{i-}S_{i+})+s_{iz}S_{iz}]$, where $s_{i\pm}=s_{ix}\pm is_{iy}$ and $S_{i\pm}=S_{ix}\pm iS_{iy}$. At $T=0$, the localized $d$ spin has the largest $S_{iz}$ of $S_{iz}=2$. Thus, the term $s_{i+}S_{i-}$ in the $s$-$d$ exchange interaction provides a decrease in $S_{iz}$ of the localized $d$ spin and an increase in $s_{iz}$ of conduction $s$ electrons, namely, down-to-up spin-flip $s$-electron scattering represented by ${\rm Im}(\Sigma_{s\downarrow})$. This is the reason for the relation $|{\rm Im}(\Sigma_{s\downarrow})| \gg |{\rm Im}(\Sigma_{s\uparrow})|$ at low temperature ($T<100\,\,{\rm K}$). When the temperature is increased over 100\,\,K, the localized $d$ spin comes to have a smaller $S_{iz}$, which enhances the up-to-down $s$-electron scattering through the term $s_{i-}S_{i+}$. This also provides a saturation of the down-to-up $s$-electron scattering, since the effect of the term $s_{i+}S_{i-}$ is relatively weakened. These are characterized by an increase in $|{\rm Im}(\Sigma_{s\uparrow})|$ [Fig. \ref{Fig2}(b)] and a saturation of $|{\rm Im}(\Sigma_{s\downarrow})|$ [Fig. \ref{Fig2}(c)], respectively. In this work, we neglected the $p$-$d$ exchange interaction as mentioned above, since the energy levels of $p$ states in Fe are much higher than $E_{\rm F}$ ($E-E_{\rm F}\approx1\,\,{\rm eV}$) and the $p$-$d$ exchange interaction has little effect on our results. We confirmed this point by similar calculations including the $p$-$d$ exchange interaction (see the Supplemental Material \cite{Supplemental}).

Let us further discuss the reduction of the TMR ratio from the viewpoint of electronic structures. Figure \ref{Fig3}(a) shows the conductances and TMR ratio as a function of temperature at $J_{sd}=-1.5\,\,{\rm eV}$. As is shown later, this value of $J_{sd}$ is close to the one estimated by a non-empirical method. When the temperature increases, the antiparallel conductance $G_{\rm AP}=G_{{\rm AP},\uparrow}+G_{{\rm AP},\downarrow}$ largely increases ($G_{\rm AP}$ at 300\,\,K is almost twice as large as that at 10\,K) while the parallel conductance $G_{\rm P}=G_{{\rm P},\uparrow}+G_{{\rm P},\downarrow}$ hardly changes \cite{remark_conductance}, leading to the significant reduction of the TMR ratio. Such a dominance of $G_{\rm AP}$ in the temperature dependence of the TMR ratio is consistent with experimental results in various MTJs \cite{2009Ishikawa-APL,2021Scheike-APL,2016Belmoubarik-APL}. To deeply understand this behavior, we next focus on the ${\bf k}_\parallel$-resolved conductances shown in Figs. \ref{Fig3}(d)--\ref{Fig3}(l). At low temperature ($T=10\,\,{\rm K}$), the well-known features of the $\Delta_1$ coherent tunneling \cite{2001Butler-PRB,2001Mathon-PRB} are seen; the up-spin conductance $G_{{\rm P},\uparrow}({\bf k}_\parallel)$ in the parallel magnetization state [Fig. \ref{Fig3}(d)] has a broad peak centered at ${\bf k}_\parallel=(0,0)=\Gamma$ while the down-spin one $G_{{\rm P},\downarrow}({\bf k}_\parallel)$ [Fig. \ref{Fig3}(g)], which has only a small value at the $\Gamma$ point, instead has relatively large values in a ring-shaped region surrounding the $\Gamma$ point. Such a clear difference in the conductance can be naturally explained by the half metallicity in the $\Delta_1$ state of Fe [Figs. \ref{Fig3}(b) and \ref{Fig3}(c)], as shown by pioneering theoretical studies \cite{2001Butler-PRB,2001Mathon-PRB}. In the antiparallel magnetization state [Fig. \ref{Fig3}(j)], the conductance has only a very small value at the $\Gamma$ point owing to the absence of the $\Delta_1$ down-spin band crossing $E=E_{\rm F}$ [Fig. \ref{Fig3}(c)]. When we increase the temperature to 50\,\,K, the effect of the spin mixing clearly appears: The down-spin conductance $G_{{\rm P},\downarrow}({\bf k}_\parallel)$ [Fig. \ref{Fig3}(h)] has large values around the $\Gamma$ point and the value just at the $\Gamma$ point is not so small, which comes from the feature of the up-spin conductance $G_{{\rm P},\uparrow}({\bf k}_\parallel)$. On the other hand, the values of $G_{{\rm P},\uparrow}({\bf k}_\parallel)$ hardly changes [Fig. \ref{Fig3}(e)], since the spin-mixing effect from $G_{{\rm P},\downarrow}({\bf k}_\parallel)$ is quite small owing to the relation $G_{{\rm P},\downarrow}({\bf k}_\parallel) \ll G_{{\rm P},\uparrow}({\bf k}_\parallel)$. In the antiparallel state [Fig. \ref{Fig3}(k)], the conductance at the $\Gamma$ point largely increases compared to Fig. \ref{Fig3}(j) due to the spin-mixing effect in the $\Delta_1$ state, which is the reason why $G_{{\rm AP},\uparrow}\,(=G_{{\rm AP},\downarrow})$ significantly increases as shown in Fig. \ref{Fig3}(a). In other words, $\Delta_1$ electrons scattered from the up-spin to down-spin state contribute dominantly to the enhancement of $G_{\rm AP}$ and thereby the reduction of the TMR ratio. When we increase the temperature to 100\,K, $G_{{\rm P},\downarrow}({\bf k}_\parallel)$ and $G_{{\rm AP},\uparrow}({\bf k}_\parallel)$ [Figs. \ref{Fig3}(i) and \ref{Fig3}(l)] increase further while $G_{{\rm P},\uparrow}({\bf k}_\parallel)$ [Fig. \ref{Fig3}(f)] hardly changes, leading to a further reduction of the TMR ratio.

\begin{figure}
\includegraphics[width=8.1cm]{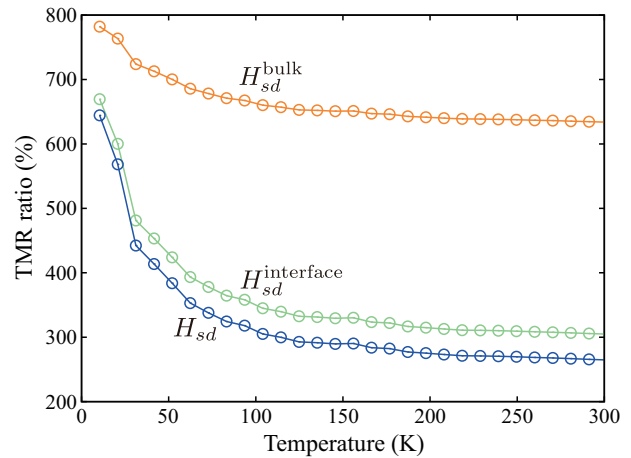}
\caption{\label{Fig4} Temperature dependencies of the TMR ratio calculated with $J_{sd}=-1.5\,{\rm eV}$ for three different cases: The $s$-$d$ exchange interaction was considered in the whole region of the electrodes ($H_{sd}$), only in the interface region ($H^{\rm interface}_{sd}$), and only in the bulk region ($H^{\rm bulk}_{sd}$).}
\end{figure}
To see the effect of spin mixing at different regions of ferromagnetic layers, we calculated the TMR ratio for two additional cases: (i) The $s$-$d$ exchange interaction was considered only in the interfacial Fe layers ($H^{\rm interface}_{sd}$ in Fig. \ref{Fig4}); and (ii) the $s$-$d$ exchange interaction was considered only in the bulk regions of Fe except the interfacial Fe layers ($H^{\rm bulk}_{sd}$ in Fig. \ref{Fig4}). Figure \ref{Fig4} shows the temperature dependencies of the TMR ratio for the three cases, which clarifies that the spin mixing at interfacial Fe layers provides the dominant contribution to the sharp reduction of the TMR ratio. This indicates the importance of selecting ferromagnets with small $|J_{sd}|$ at the interface for preventing the temperature decay of the TMR ratio.

Finally, we estimate the coupling constant $J_{sd}$ using a nonempirical method. A pioneering theory by Schrieffer and Wolff \cite{1966Schrieffer-PR} has shown that the $s$-$d$ exchange interaction [Eq. (\ref{eq2})] can be derived by applying a canonical transformation to the Anderson Hamiltonian and the coupling constant $J_{sd}$ can be expressed as
\begin{equation}
J_{sd} \approx |V_{sd}|^2 \frac{U}{\epsilon_d\,(\epsilon_d+U)},\label{eq3}
\end{equation}
where $\epsilon_d$ is the energy level of a $d$ orbital with respect to $E_{\rm F}$, $U$ is the Coulomb interaction in the $d$ orbital, and $V_{sd}$ is the hybridization between the $d$ and $s$ states. This expression can be understood from virtual electron transitions in the second-order perturbation processes \cite{1996Yosida-Springer}. Here, we apply Eq. (\ref{eq3}) to bcc Fe$_{1-x}$Co$_x$ ($0\! \leq\! x\! \leq\! 1$), since this series of materials is typically used for MTJs with an MgO tunnel barrier and is known to give high TMR ratios \cite{2004Parkin-NatMat,2004Yuasa-NatMat,2006Yuasa-APL}. 

The values of $\epsilon_d$, $U$, and $V_{sd}$ can be estimated on the basis of the maximally localized Wannier function (MLWF) method implemented in the {\scriptsize RESPACK} code \cite{2021Nakamura-CPC}. We first conducted density functional theory (DFT) calculations of bcc Fe$_{1-x}$Co$_x$ using the {\scriptsize QUANTUM ESPRESSO} code \cite{2009Giannozzi-JPCM}. We employed the Perdew--Burke--Ernzerhof exchange-correlation potential \cite{1996Perdew-PRL} and the optimized norm-conserving Vanderbilt (ONCV) pseudopotentials from PseudoDojo \cite{2018Setten-CPC}. The primitive bcc unit cell with a lattice parameter of 2.866\,\AA\, was used for all $x$. The alloys with $0 < x < 1$ were treated by the virtual crystal approximation. We used $10\times10\times10$ {\bf k}-point grids and an energy cutoff of 108\,\,Ry for the wave functions and 432\,\,Ry for the electron charge densities. We next constructed the MLWFs \cite{1997Marzari-PRB,2001Souza-PRB} using the {\footnotesize RESPACK} code \cite{2021Nakamura-CPC}. By adopting atomic $s$, $p$, and $d$ orbitals as initial projection functions, we obtained nine MLWFs that reproduce the original DFT band dispersion around the Fermi level. Here, the inner and outer energy windows were set to [1\,\,eV,\,30\,\,eV] and [0\,\,eV,\,\,55\,eV], respectively, for all $x$, where the Fermi level was located at 17.86\,\,eV for $x=0$ (Fe) and 17.17\,\,eV for $x=1$ (Co). The obtained MLWFs are not the same as the initial atomic orbitals but sufficiently maintain their features. We also obtained the hopping integrals and the on-site energy of each MLWF. We used the nearest-neighbor hopping integrals between $s$ and $d$ orbitals as $V_{sd}$ and on-site energies of $d$ orbitals as $\epsilon_d$. The Coulomb interaction parameters were also calculated using the {\footnotesize RESPACK} code. Here, we adopted the usual random phase approximation for calculating the dielectric function \cite{2021Nakamura-CPC}. The energy cutoff for the dielectric function was set to 60\,Ry for ensuring good convergence of the Coulomb interaction parameters. The polarization function was calculated using 60 bands. The obtained intraorbital screened Coulomb interaction $U$ in each $d$ orbital was used to estimate $J_{sd}$ given by Eq. (\ref{eq3}). The screened Hund exchange interaction between $s$ and $d$ electrons was also obtained for each $d$ orbital.

\begin{figure}
\includegraphics[width=8.7cm]{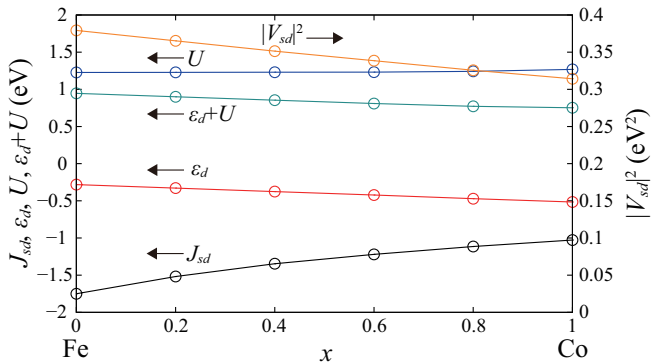}
\caption{\label{Fig5} The $x$ dependencies of $J_{sd}$, $\epsilon_d$, $U$, $\epsilon_d+U$, and $|V_{sd}|^2$ in bcc Fe$_{1-x}$Co$_x$.}
\end{figure}

Figure \ref{Fig5} shows $x$ dependencies of $J_{sd}$, $\epsilon_d$, $U$, and $|V_{sd}|^2$ averaged over the $d$ orbitals. First of all, $J_{sd}$ is negative (i.e., antiferromagnetic coupling) for all the values of $x$, since $\epsilon_d$ and $\epsilon_d+U$ have different signs. As $x$ is increased from 0 (Fe) to 1 (Co), the $d$ level $\epsilon_d$ gets deeper, which is natural since Co has more valence electrons than Fe. This also decreases the hybridization $|V_{sd}|$ between the $d$ state and the $s$ state near $E_{\rm F}$. These changes in $\epsilon_d$ and $|V_{sd}|$ lead to a decrease in $|J_{sd}|$ with increasing $x$. We summarized the values of $J_{sd}$ and $\Delta {\rm TMR}$ in Table \ref{tab}, indicating that the reduction in the TMR ratio monotonously decreases with increasing $x$. This tendency is consistent with previous experimental results on Fe/MgO/Fe ($\Delta {\rm TMR}\sim -500\%$) \cite{2021Scheike-APL} and Co/MgO/Co ($\Delta {\rm TMR}\,\, {\small \lesssim} -100\%$) \cite{2006Yuasa-APL}.

\begin{table}
\caption{\label{tab}
The $x$ dependencies of $J_{sd}$ and the reduction of the TMR ratio at room temperature, $\Delta {\rm TMR}\!=\,$TMR ratio (300\,\,K)\,$-$\,TMR ratio (0\,\,K), in bcc Fe$_{1-x}$Co$_x$ \cite{remark_TMR-CoFe}.
}
\begin{ruledtabular}
\begin{tabular}{ccccccc}
$x$ & 0 & 0.2 & 0.4 & 0.6 & 0.8 & 1.0 \\
\hline
$J_{sd}\,({\rm eV})$ & --1.75 & --1.52 & --1.35 & --1.22 & --1.11 & --1.03 \\
$\Delta {\rm TMR}$\,(\%)  & --610 & --560 & --510 & --460 & --420 & --380 \\
\end{tabular}
\end{ruledtabular}
\end{table}
Note that there exists another contribution to $J_{sd}$ different from Eq. (\ref{eq3}). It is the Hund exchange interaction \cite{1983Oles-PRB}, more generally called the direct exchange interaction \cite{1996Yosida-Springer}, between $s$ and $d$ electrons. Using the MLWF method, we estimated its contribution to $J_{sd}$ and obtained small values of 0.25--0.28\,eV for all the values of $x$ in Fe$_{1-x}$Co$_x$. Therefore, we can conclude that the $s$-$d$ exchange interaction due to the second-order perturbation [Eq. (\ref{eq3})] provides the dominant contribution to $J_{sd}$. The total value of $J_{sd}$ including both the contributions is estimated to be $\sim -1.5\,{\rm eV}$ for $x=0$ (Fe), which justifies our choice of $J_{sd}=-1.5\,{\rm eV}$ in Figs. \ref{Fig3} and \ref{Fig4}.

In summary, we theoretically investigated the temperature dependence of the TMR effect in Fe/MgO/Fe(001). We clarified a crucial importance of the $s$-$d$ exchange interaction in the degradation of the TMR at finite temperature: The $s$-$d$ exchange interaction at the interfacial ferromagnetic layers provides spin-flip scattering in the $\Delta_1$ states, leading to a significant reduction of the TMR ratio. To the best of our knowledge, most of the previous theories on the TMR effect might have missed this fact, since they have focused only on the $d$-$d$ exchange interaction in bulk ferromagnets. Our findings are also supported by the experimental fact that conduction $sp$-electron states provide dominant contributions to the transport properties in MTJs. We finally estimated the coupling constant $J_{sd}$ of the $s$-$d$ exchange interaction on the basis of a non-empirical method. By using the present approach, one can predict the material dependence of the TMR reduction at room temperature for a wide range of ferromagnets, which would be quite useful for designing MTJs with a weak temperature dependence of the TMR ratio.

The authors are grateful to H. Itoh, S. Honda, M. Matsumoto, and H. Sukegawa for helpful discussions. This work was partly supported by TDK Corporation and Grant-in-Aids for Scientific Research (Grants No. JP16H06332, No. JP17H06152, No. JP20H02190, and No. JP20K14782). The crystal structures were visualized using {\footnotesize VESTA} \cite{2011Momma-JAC}.


\begin{thebibliography}{99}
\bibitem{remark_kpara-dep} We focused on the central region of the first Brillouin zone, since each conductance does not have significant values outside of this region.
\bibitem{remark_conductance} Note that $\Gamma_{{\rm P},\uparrow}$ is more than one order larger than $\Gamma_{{\rm P},\downarrow}$ and thus $\Gamma_{\rm P} \approx \Gamma_{{\rm P},\uparrow}$.
\end{thebibliography}
\end{document}